\documentclass[12pt]{article}
\usepackage{amsmath,bm}
\usepackage{amssymb}
\usepackage{graphicx}
\usepackage{amsfonts}         
\usepackage{fancybox}   
\usepackage{yfonts} 
\usepackage{hyperref}

\renewcommand{\baselinestretch}{1.4}
\def\comments#1{}

\def\Re{{\rm Re\hskip0.1em}}
\def\Im{{\rm Im\hskip0.1em}}

\def\vev#1{\langle{#1}\rangle}

\def\CC{{\cal C}}

\def\CE{{\cal E}}
\def\CF{{\cal F}}

\def\CN{{\cal N}}
\def\CO{{\cal O}}
\def\CP{{\cal P}}
\def\CL{{\cal L}}

\def\II{\relax{I\kern-.10em I}}

\def\IB{\relax{\rm I\kern-.18em B}}
\def\ID{\relax{\rm I\kern-.18em D}}
\def\IE{\relax{\rm I\kern-.18em E}}
\def\IF{\relax{\rm I\kern-.18em F}}
\def\IG{\relax\hbox{$\inbar\kern-.3em{\rm G}$}}
\def\IGa{\relax\hbox{${\rm I}\kern-.18em\Gamma$}}
\def\II{\relax{\rm I\kern-.18em I}}
\def\IK{\relax{\rm I\kern-.18em K}}

\def\inbar{\,\vrule height1.5ex width.4pt depth0pt}

\def\frac#1#2{{#1 \over #2}}

\newdimen\tableauside\tableauside=1.0ex
\newdimen\tableaurule\tableaurule=0.4pt
\newdimen\tableaustep
\def\phantomhrule#1{\hbox{\vbox to0pt{\hrule height\tableaurule width#1\vss}}}
\def\phantomvrule#1{\vbox{\hbox to0pt{\vrule width\tableaurule height#1\hss}}}
\def\sqr{\vbox{%
  \phantomhrule\tableaustep
  \hbox{\phantomvrule\tableaustep\kern\tableaustep\phantomvrule\tableaustep}%
  \hbox{\vbox{\phantomhrule\tableauside}\kern-\tableaurule}}}
\def\squares#1{\hbox{\count0=#1\noindent\loop\sqr
  \advance\count0 by-1 \ifnum\count0>0\repeat}}
\def\tableau#1{\vcenter{\offinterlineskip
  \tableaustep=\tableauside\advance\tableaustep by-\tableaurule
  \kern\normallineskip\hbox
    {\kern\normallineskip\vbox
      {\gettableau#1 0 }%
     \kern\normallineskip\kern\tableaurule}%
  \kern\normallineskip\kern\tableaurule}}
\def\gettableau#1 {\ifnum#1=0\let\next=\null\else
  \squares{#1}\let\next=\gettableau\fi\next}

 \def\eqnn#1{\xdef #1{(\secsym\the\meqno)}\writedef{#1\leftbracket#1}%
 \global\advance\meqno by1\wrlabeL#1}
 \def\eqna#1{\xdef #1##1{\hbox{$(\secsym\the\meqno##1)$}}
 \writedef{#1\numbersign1\leftbracket#1{\numbersign1}}%
 \global\advance\meqno by1\wrlabeL{#1$\{\}$}}
 \def\eqn#1#2{\xdef #1{(\secsym\the\meqno)}\writedef{#1\leftbracket#1}%
 \global\advance\meqno by1$$#2\eqno#1\eqlabeL#1$$}

\global\newcount\itemno \global\itemno=0

\def\itemaut#1{\global\advance\itemno by1\noindent\item{\the\itemno.}#1}

\def\del{\partial}

\def\({\left(}
\def\){\right)}
\def\kkk{{\mathfrak{K}}}

\def\eg{{\it e.g.}}
\def\ie{{\it i.e.}}

\newif{\ifeq}           
\eqtrue                 
\newcommand{\be}{\begin{equation}}
\newcommand{\ee}{\end{equation}}
\newcommand{\bea}{\begin{eqnarray}}
\newcommand{\eea}{\end{eqnarray}}
\newcommand{\bean}{\begin{eqnarray*}}
\newcommand{\eean}{\end{eqnarray*}}

\def\({\left(}
\def\){\right)}
\def\[{\left[}
\def\]{\right]}

\newcommand{\half}{\frac{1}{2}}

\renewcommand{\O}{{\cal O}}

\def\CO{\O}

\newcommand{\IR}{{\mathbb R}}
\newcommand{\IZ}{{\mathbb Z}}

\def\ie{{\it i.e.}}

\newcommand{\lsim}{\,\raise.3ex\hbox{$<$\kern-.75em\lower1ex\hbox{$\sim$}}\,}
\newcommand{\gsim}{\,\raise.3ex\hbox{$>$\kern-.75em\lower1ex\hbox{$\sim$}}\,}

\newif{\ifeq}
\eqtrue

\def\vx{\vec{x}}

\def\mass{\ell}

\numberwithin{equation}{section}

\global\newcount\itemno \global\itemno=0

\def\itemaut#1{\global\advance\itemno by1\noindent\item{\the\itemno.}#1}

\begin{document}

\begin{titlepage}

\begin{flushright}
MIT-CTP/4070, NSF-KITP-09-137
\end{flushright}
\vfil

\begin{center}
{\huge 
The particle number 
in Galilean holography}\\
\end{center}
\vfil
\begin{center}
{\large Koushik Balasubramanian and John McGreevy}\\
\vspace{1mm}
Center for Theoretical Physics, MIT,
Cambridge, Massachusetts 02139, USA\\
{\tt koushikb at mit.edu}\\
\vspace{3mm}
\end{center}

\begin{center}
{\large Abstract}
\end{center}
\noindent
Recently, gravity duals for certain Galilean-invariant conformal field theories have been constructed.
In this paper, 
we point out that the spectrum of the particle number operator
in the examples found so far is not 
a necessary consequence of the existence of a gravity dual.
We record some progress towards more realistic spectra.
In particular, we construct bulk systems with asymptotic
Schr\"odinger symmetry and only one extra dimension.
In examples, we find solutions 
which describe these
 Schr\"odinger-symmetric systems at finite density.
A lift to M-theory is used to resolve a curvature singularity.
As a happy byproduct of this analysis, we realize a state which 
could be called a holographic Mott insulator.


\vfill
\begin{flushleft}
July 2010
\end{flushleft}
\vfil
\end{titlepage}
\newpage
\renewcommand{\baselinestretch}{1.1}  

\renewcommand{\arraystretch}{1.5}

\tableofcontents

\section{Introduction}

Particle production is a dramatic, necessary consequence of relativistic field theory.
There is no particle production in Galilean-invariant field theories, which 
therefore have an
extra conserved quantity.
This quantity is often thought of as the particle number, 
but (since we can and will work in units of the mass of one particle)
it is equivalent to the total rest mass.
In systems with multiple species, to be discussed more below,
the latter definition is the more useful one.

Recently, candidate gravity dual descriptions of 
certain Galilean-invariant, scale-invariant 
field theories were proposed \cite{Son:2008ye, Balasubramanian:2008dm}.
More specifically, these theories are non-relativistic conformal field theories (NRCFTs):
their symmetry group, the Schr\"odinger group, contains a special conformal generator.
This same group is respected by the dynamics of fermions with unitarity-limited two-body interactions \cite{Mehen:1999nd, Nishida:2007pj},
which arise by tuning ultracold fermionic atoms to a Feshbach resonance.
These gravity solutions have since been embedded in string theory
and put at finite temperature and density 
\cite{Maldacena:2008wh, Herzog:2008wg, Adams:2008wt, Kovtun:2008qy, Hartnoll:2008rs}.\footnote{Related earlier
work on geometric realizations of the Schr\"odinger group 
includes \cite{bateman, Duval, Horvathy}.
Earlier work on holography for spaces with degerate boundaries includes
\cite{BrittoPacumio:1999sn, Taylor:2000xf}.
Subsequent work, including examples of Schrodinger-invariant supergravity solutions which preserve some supersymmetry, includes \cite{Yamada:2008if, Colgain:2009wm}.}

In the known examples, the symmetry associated
with the conserved rest mass is realized geometrically in the gravity 
dual as the isometry of a circle, whose coordinate we call $\xi$.  
Compactifying on a circle with circumference $L_\xi$ produces
a spectrum of possible values of the rest mass of states in the theory of the form
\be\label{massspectrum} \{ {\rm masses}\} = {1\over 2\pi L_\xi}~\IZ_+ ~~. \ee
The main purpose of this (somewhat polemical) article is to point out that this particular spectrum is
not a necessary consequence of the existence of a gravity dual.

The form of the spectrum \eqref{massspectrum}
seems to be responsible for the strange thermodynamics 
found in \cite{Maldacena:2008wh, Herzog:2008wg, Adams:2008wt, Kovtun:2008qy}:
\be\label{DLCQthermo} F \sim  -{T^4 \over \mu^2}, ~ \mu < 0 . \ee
This is quite different from the behavior of unitary fermions, where
in particular the chemical potential is positive, and the free energy scales
like a positive power of $\mu$.
These theories are closely related to relativistic field theories,
via (modifications of) the discrete lightcone quantization procedure (DLCQ).
This fact is made particularly vivid in the calculation of 
the free energy 
\eqref{DLCQthermo} from a free relativistic field theory
in DLCQ by \cite{fuertesDLCQ}.
The modifications of DLCQ in \cite{nullmelvin} 
(associated with ``$\beta$-deformation")
simplify the theory by 
removing most of the troublesome \cite{Hellerman:1997yu} lightcone zeromodes,
but do not change the spectrum of the lightcone momentum operator, $i \partial_\xi$.

In this work, our goal is to learn how to construct gravity duals for
NRCFTs 
with other (ideally, more realistic) spectra.
We demonstrate that it is 
not necessary to realize the Schr\"odinger algebra 
in a gravity dual entirely via isometries of the bulk metric.
It was natural to try to realize the full
algebra by isometries, since 
the obviously-geometric momentum and boost generators
commute to the particle number operator $\hat N$, 
\be\label{particlenumberalg}[K_i, P^j] = iN \delta^i_j . \ee
However, here we show that 
this algebra can be realized without the introduction of a $\xi$ dimension, 
if the boost generator
acts by gauge transformations on fields charged
under an additional abelian gauge symmetry in the bulk.
The construction is quite similar to the way that 
these symmetries are realized on states 
of a quantum system \cite{inonu}: a Galilean boost by velocity $\vec v$ acts on 
the phase of the wavefunction (in the Schr\"odinger representation) of a particle of mass $m$ by
\be  \psi(x,t)\mapsto e^{ i m \( \half v^2 t + \vec v \cdot \vec x \) } \psi(x-vt,t) ~; \label{inonu}\ee
from this expression one can show that (\ref{particlenumberalg}) is satisfied.

Using this idea we construct solutions describing $d=2,~ z=2$ NRCFT without the additional circle. 
For most of this paper we will 
employ a practical approach to holography
advocated in \eg\ \cite{Hartnoll:2009qx, McGreevy:2009xe}: 
we do not yet know the constraints 
that quantum gravity imposes
on effective field theories of gravity coupled to matter
(known examples \cite{wittenbagger, AlvarezGaume:1983ig, hellerman} are not very forceful, 
and it is clear that our grasp on the space of less-supersymmetric string vacua is poor),
and so we will employ the simplest gravity models 
with which we can approach the physics of interest.
We will, however, find it useful in \S\ref{sec:mott} to 
lift one of our solutions to 11-dimensional supergravity
in order to resolve a curvature singularity.
That solution describes a system (at finite density and at zero temperature) with a gap for the charged excitations;
it appears to provide a holographic description of a Mott insulator.
This is an improvement over a previous holographic realization of an insulating state \cite{Nishioka:2009zj},
which had zero density.
We also succeed in constructing some examples
where there are several species of particles,
so that the spectrum of the number operator is not just
integer multiples of a single mass;
this is described in \S\ref{sec:multiplespecies}.
Finally, in \S\ref{sec:SF},
we discuss a conjugate issue, namely
whether the NRCFTs described by the constructions of \cite{Son:2008ye, Balasubramanian:2008dm}
have superfluid groundstates (at low temperature and finite density), and what 
such a state would look like from the point of view of the gravity dual.
We relegate to an appendix a curious black hole solution with the new realization of asymptotic Schr\"odinger symmetry.

\section{Getting rid of the $\xi$ direction}

In this section, we study the dimensional reduction
on the particle-number circle
of the systems discussed previously in \cite{Son:2008ye, Balasubramanian:2008dm}.
We are doing this because it provides a proof of principle
that there can be gravity theories with Schr\"odinger symmetry
which don't have this annoying extra dimension.
Our real goal is to find new solutions where the
spectrum of the mass operator can be different (\ie\ 
not the KK tower of momentum modes on a circle),
and where the thermodynamics may therefore
be more like that of unitary fermions.
Our immediate goal is to understand how the symmetries are realized.

A concern which remains even after dimensionally reducing to
replace the role of the $\xi$-dimension with a 
gauge field in a lower-dimensional description
is charge-conjugation invariance.
In a relativistic QFT with charge-conjugation invariance (like the 
one living in the bulk here), 
the spectrum of a $U(1)$ symmetry must include both positive and negative charges.
Below, we explicitly break this symmetry by imposing boundary conditions
which introduce a background electric field.

\subsection{Review of gravity duals with Schr\"odinger symmetry}

In this paper, we are interested in 
non-relativistic conformal field theories which are governed by the Schr\"odinger symmetry algebra. The Schr\"odinger group includes Galilean invariance, scale invariance and one special conformal transformation. Geometries which realize these symmetries as isometries were constructed in \cite{Son:2008ye, Balasubramanian:2008dm} and the metric is
\be
\label{eq:schrodmetric} 
ds^{2 } = -\alpha^{2} {dt^{2}\over r^{4}} + {2d\xi dt + d\vec{x}^{2} + dr^{2} \over r^{2}} L_{{\rm AdS}}^2 
\ee
where $\vec{x}$ is a vector of $d$ spatial dimensions,
and we will work in units with $L_{{\rm AdS}} = 1$. This metric solves the equations of motion of Einstein gravity coupled to a massive gauge field and a negative cosmological constant. 
In the above metric, $\xi$ is a compact direction and the particle number symmetry is realized as translation symmetry along this circle. 
When $\alpha$ is zero, the metric is just $AdS$ in light cone coordinates with one of the null directions compactified \cite{Goldberger:2008vg, Barbon:2008bg}.

This solution can be embedded in string theory \cite{Maldacena:2008wh, Herzog:2008wg, Adams:2008wt} and the dual field theory is a modified DLCQ of $\CN=4$ SYM theory (or of another quiver gauge theory 
dual to type IIB on $AdS_5$ times a Sasaki-Einstein manifold). 
A black hole solution 
asymptotic to the metric (\ref{eq:schrodmetric})
was found in \cite{Maldacena:2008wh, Herzog:2008wg, Adams:2008wt}; this 
describes the dual NRCFT at finite density and finite temperature. 
In the black hole solution, $\xi$ is not null 
everywhere
because $g_{\xi\xi}$ is not identically zero as in the vacuum solution.
This implies that the radius of the circle is non-zero in the bulk and the supergravity approximation can be trusted in regions where the radius is large compared to the string length scale. Thus the nonzero $g_{\xi \xi}$ component acts like a regulator and this fact will be used here to construct alternate holographic descriptions of Schr\"odinger algebra.   

Dimensional reduction of this solution 
along $\xi$ yields a lower-dimensional system with asymptotic Schr\"odinger symmetries. The matter content of the lower-dimensional gravity theory consists of a massive vector field, $U(1)$ gauge field and two scalars (higher-dimensional dilaton and the radion). We would like to have a simpler system that can aid us in understanding the lower dimensional realization of Schr\"odinger symmetry. 

In \S3.4 of \cite{Adams:2008wt}, we studied a scaling limit of this black hole solution (finite $\mu, T$ )
which had zero temperature, but had a non-trivial $g_{\xi\xi}$ component 
(let's call this solution $Sch_{\Omega\neq0, T=0}$). 
This solution is singular in the IR ($r \to \infty$) and should
not be taken too seriously.  
We will use it here as a helpful device to 
learn about possible bulk realizations of the Schr\"odinger algebra.

\subsection{Dimensional Reduction of $Sch_{\Omega\neq0, T=0}$}

The geometry of $Sch_{\Omega\neq0, T=0}$ is described by the following line element
\be \label{gHd} ds^2_{5} = {1\over r^2 \kappa^{2/3}}\( {-dt^2\over r^2} + 2 dtd\xi + (\kappa-1) r^2 d\xi^2\) + \kappa^{1/3} \({d\vx^2 + dr^2 \over r^2}\)\ee
where $\kappa = 1 + \Omega^2 r^2$, for some constant $\Omega$ which determines the density.
This can be obtained as a classical solution of the following action\footnote
{
The potential $V(\Phi)$ is \cite{Maldacena:2008wh, Herzog:2008wg}
$V(\Phi) = 4 e^{2 \Phi/3} \( e^{2 \Phi} - 4 \) $
but it will disappear soon.
}:
\be \label{actionHd} S_{5} = {C_0} \int d^{5}x\sqrt{-g_{5}}\[R_{5} - {4\over 3} \(\nabla \Phi\)^2 + V\(\Phi\) - {1\over 4} e^{-8\Phi/3} F_{5}^2 -{m^2\over2}A_{5}^2\]\ee 
with $\Phi = -\half \log \kappa$, $A_{5}= \half r^{-2}\kappa^{-1}\(dt + (\kappa-1)r^2d\xi\)$ and $F_{5}=dA_{5}$.
In this subsection, we will perform a series of manipulations using this solution 
to identify a $4$-dimensional system that admits 
asymptotic solutions which respect the Schr\"odinger group in two space dimensions.

Dimensional reduction of the above action along $\xi$ direction yields a lower-dimensional system with 
Schr\"odinger symmetry. In this system, the particle number symmetry is realized as a bulk gauge symmetry. The $g_{\xi\xi}$ component of the higher-dimensional metric appears as a scalar field (the radion field $e^{2\sigma}$) in the lower-dimensional system. 
With the metric ansatz 
\be
 ds^2_{5} =  ds^2_4 +  e^{2\tilde \sigma}\( d\xi + {B}\)^2 \ee
the lower-dimensional action can be written as
$$S_{4} = {C_0}L_{\xi} \int d^{4}x~ e^{\tilde \sigma}\sqrt{-G_{D}}\Big[R_{D} - {4\over 3} \(\nabla \Phi\)^2 + V\(\Phi\) - {1\over 4} e^{-8\Phi/3} F_D^2 -{m^2\over2}A_D^2+$$
\be \label{actionLdOm}  4 \nabla \Phi\cdot \nabla \tilde \sigma -{e^{2\tilde \sigma}\over4} \(dB\)^2 
-{1\over 2}e^{-8 \Phi/3}\(\nabla A_\xi\)^2 -{m^2\over2}A_\xi^2  + \CL_{int}\( A_\xi, B,A\) \Big]~.\ee 
Note that the line element in $(\ref{gHd})$ can be written as
\be \label{gHd1} ds^2_{5} = {1\over r^2 \kappa^{2/3}}\( {-dt^2\over r^2} + \Omega^2 r^4\( d\xi + {dt\over r^4 \Omega^2}\)^2 -{dt^2 \over \Omega^2 r^4} \) + \kappa^{1/3} \({d\vx^2 + dr^2 \over r^2}\)~~.\ee
Scaling $t$ by $\Omega Q^{1/2}$ and scaling $\xi$ by $Q^{-1/2}/\Omega$   in the above expression we get
\be \label{gHd2} ds^2_{5} = {1\over r^2 \kappa^{2/3}}\( \kappa Q{-dt^2\over r^4} + r^4 / Q  \( d\xi + Q{dt\over r^4}\)^2  \) + \kappa^{1/3} \({d\vx^2 + dr^2 \over r^2}\)~~.\ee
Under this rescaling $A_t \rightarrow \Omega Q^{1/2} A_t$.
Hence, the higher-dimensional line element can be written as
 \bea
 \label{gHd3} ds^2_{5} =  ds^2_D +  e^{2 \tilde \sigma}\( d\xi + {B}\)^2  \qquad\\
 \label{sigmaANDB} e^{2\tilde \sigma} = \frac{r^2}{Q\kappa^{2/3}} , \qquad B =\frac{Q}{r^4}dt\quad.
 \eea
The $4-$dimensional line element in the above expression is
\be \label{gLd} ds^2_{D} \equiv (G_D)_{\mu\nu}dx^{\mu}dx^\nu = \kappa^{1/3} \(-{Q dt^2\over r^6} + {d\vx^2 + dr^2 \over r^2}\)  \ee
If we now 
define $e^{ 2 \sigma} = \Omega^2 e^{ 2 \tilde \sigma}$ and 
take the scaling limit, $\Omega \rightarrow 0$, holding $\sigma$ fixed,
we are left with an extremum of the much-simpler action
\be \label{actionLd}  S_{4} = {C_0} L_\xi  \int d^{4}x ~e^{\sigma}\sqrt{-G_{D}}\Big[R_{D} - 2 \Lambda  -{e^{2\sigma}\over4} \(dB\)^2  \Big]~,\ee 
where we have chosen units so that the cosmological constant is $\Lambda = - 6$.
In the above limit, $A=0$, $\Psi =0$, $\Phi =0$ and $\kappa =1$;
the $4$-dimensional solution 
is \eqref{gLd} with $\kappa=1$.
Note that $Q$ is related to the chemical potential. When $Q \to \infty$, the Schr\"odinger symmetries become an exact symmetry of the above system; however, the metric becomes degenerate in the $Q \to \infty$ limit.
Rewriting the `string frame' action (\ref{actionLd}) in 4d Einstein frame 
(and throwing away the fields $A, \Psi, \Phi$ which vanish)
we see that 
\be 
\label{4dzerotemp}
ds^2_E =e^\sigma\( -Q {dt^2 \over r^6 } + {d\vec{x}^2 + dr^2 \over r^2 }\), ~~B = Q{dt \over  r^4},~~ e^{\sigma} =   {r \over \sqrt Q}\ee
is a solution of the simple action
 \be \label{actionLdE}  S^{E}_{4} =  
 \kkk
 \int d^{4}x\sqrt{-g_{4}}\Big[R_{4} - 2 \Lambda e^{-\sigma}  -{e^{3\sigma}\over4} \(dB\)^2 - {3\over 2} \(\del \sigma \)^2   \Big]~~,\ee 
 where we have named $\kkk \equiv C_0 L_\xi$ the effective 4d coupling.
 Note that the apparent strong coupling behavior of the action for the gauge field $B$ 
 at the boundary 
 ($g_{{\rm eff}}^{-2} \sim e^{3\sigma} \sim r^3 \to 0$) is an artifact of dimensional reduction.

 \subsection{Symmetry Generators}
 \label{sec:sym}

 Let us try to understand how the Schr\"odinger symmetry group is realized by the above action
 and asymptotics. It is clear that the symmetries of the Schr\"odinger group are not realised as isometries in the lower-dimensional theory: the putative symmetry generators in the lower-dimensional theory do not solve the Killing equation.  The metric 
 is of the Lifshitz form 
 \cite{Kachru:2008yh} and 
 seems to have scaling symmetry with dynamical exponent $z=3$. 
 What equation determines the symmetry generators of the lower-dimensional action? 
 It is not hard to guess that the appropriate symmetry generators 
should solve the equation obtained by dimensional reduction of the higher-dimensional Killing equation.

So the symmetry generators of the lower-dimensional theory are:  \begin{itemize}
 \item Particle Number:  The $U(1)$ 
 gauge charge associated with the massless gauge field $B$ is the particle number.
 This acts by $B \rightarrow B + d\lambda$, and 
 by phase rotations on charged fields in the bulk, of which we should include 
 one or more.  Let us introduce such a field $\Phi \equiv |\Phi| e^{ i \varphi}$ of charge $\mass$;
 we take $\Phi$ to vanish in the solution shown above.
 $\mass$ is the mass of the associated particle.
   \item Translations and rotations are realized as usual by isometries.
 \item Galilean boosts act as follows:
 \be 
 t \to t, \quad \vec x \to \vec x - \vec v t , \quad \varphi \to
 \varphi + \mass \( \half v^2 t + \vec v \cdot \vec x \), \ee
 where $\varphi$ is the phase of a field of charge $\mass$ under 
 the particle-number gauge symmetry.  The role previously played by $\xi$
 in the Schr\"odinger geometry is now played by 
 the phase $\varphi$ of charged bulk fields.
In summary, the boost generator is:  
 $$K^{i} = -t \del_{i} \text{ + gauge shift } $$
 where the gauge transformation parameter is 
 $ \lambda =  \half v^2 t + \vec v \cdot \vec x $. 
 Note the similarity to the action in quantum mechanics 
 given in Eqn.~\eqref{inonu}.
 
   \item Scale symmetry acts by
 $$D = -2t\del_{t} - x^{i} \del_{i} -r \del_{r};$$
 \end{itemize}
The generators of these symmetries satisfy the Schr\"odinger algebra.

The asymptotic profiles of the fields are {\it not} preserved by these transformations,
but one can show, as follows, that they are nevertheless (asymptotic) symmetries of the system.
The higher dimensional (5D) Killing equation can be written as
\be \delta_{\eta} \(G_D\)_{AB} = {\CL}_{\eta} \(G_D\)_{AB} = 0  ~. \label{Killing} \ee
If the above equation is only true as $r\to 0$ (which we denote by $\approx 0$), as is the case in the solutions described above, then the symmetry is realized only asymptotically.
The lower dimensional (4D) metric ($g$) does not satisfy the 4d Killing equation, \ie\
$\delta_{\eta} g_{\mu \nu} \neq 0$.
The above equation \eqref{Killing}, however, implies\footnote{The  transformation rules for the lower-dimensional fields can be obtained from the transformation rule for the higher-dimensional metric:
\be\delta_\eta g_{\mu\nu} = \[g_{\mu\rho} \del_{\nu}\eta^{\rho} +g_{\rho\nu} \del_{\mu}\eta^{\rho}+ \eta^{\rho}\del_{\rho}g_{\mu\nu} \] \ee
\be\delta_\eta B_{\mu} = \[B_{\rho} \del_{\mu}\eta^{\rho} + \eta^{\rho}\del_{\rho}B_{\mu} \] + \del_{\mu}\eta^{D+1}\ee
\be\delta_\eta{e^{2\sigma}} = \[\eta^{\rho}\del_{\rho}e^{2\sigma}\] .\ee
Note that the quantities within $\[~\]$ are the changes due to the coordinate transformations in the lower dimensions, while the transformation of $x^{D+1}$  generates field transformations. 
}
\be \delta_{\eta} \(e^{-\sigma}g_{\mu\nu} + e^{2\sigma} B_{\mu} B_{\nu}\)\approx 0\ee
\be \delta_{\eta} \(e^{2\sigma} B_{\mu} \) \approx 0 \ee
\be \delta_{\eta} \(e^{2\sigma}\)\approx 0 \ee
These quantities have the transformation properties of tensors. We also know
(from its higher-dimensional origin) that the action can be written as a functional of these quantities, that is 
\be
S_D[g, B, \sigma] = S[e^{-\sigma} g_{\mu\nu} + e^{2\sigma} B_{\mu} B_{\nu}, e^{2\sigma} B_{\mu}, e^{2\sigma} ] ~
.\ee
 When the symmetries are realized as isometries, $\delta_\eta S$ vanishes as a consequence of 
 $\delta_\eta G_{AB}$ vanishing. In the present case, $\delta_\eta S$ will vanish as a consequence of $\delta_{\eta} \(e^{-\sigma}g_{\mu\nu} + e^{2\sigma} B_{\mu} B_{\nu}\) $, $\delta_{\eta} \(e^{2\sigma} B_{\mu} \) $ and $\delta_{\eta} \(e^{2\sigma}\)$ vanishing. 

In the solutions described above, these quantities only vanish asymptotically near the boundary.
Note that we do not know a solution of the lower-dimensional system \eqref{actionLdE}
which exactly preserves the Schr\"odinger symmetry.  This is perhaps unsurprising
given that such a solution would correspond to the vacuum of a Galilean-invariant
system, a very boring state indeed.  Rather, the surprising
fact is that the previous holographic realizations \cite{Son:2008ye, Balasubramanian:2008dm} did provide such a solution.

From the form of $B$ in \eqref{4dzerotemp}, we see that the solution (\ref{gLd}) has non-zero chemical potential ($\mu \neq 0$), but charge density zero.  
This can happen for example if the chemical potential is smaller than the particle mass.

\subsection{Wave equation}
The wave equation for a probe scalar field with charge $\mass$ under the Kaluza Klein gauge field (and mass $m^2$) takes the following form
\be \( -\omega^2 r^6+ m^2 + r^2(2 \mass \omega + k^2) \)  \Phi - r^{d+3}\del_r\(r^{-d-1} \del_r \Phi\) =0~. \ee
Notice that the first term in this equation is unimportant for the boundary behavior ($r\to 0$),
but does spoil the Schr\"odinger invariance of the equation
(and renders us and Mathematica unable to solve it analytically).

The $D$-dimensional (Einstein-frame) action that produces this equation of motion is perhaps
surprising:
\be
S_{{\rm probe}}[\Phi]
= \int \sqrt g\[ | (\partial - i \mass B ) \Phi |^2 - \( \mass^2 e^{ - 3 \sigma} + m^2 e^{ - \sigma} \)  | \Phi|^2  \]~~~~.
\ee
This coupling to the background scalar $\sigma$ is 
required in order that solutions for $\Phi$ represent 
the Schr\"odinger symmetry.

\section{Black hole solution}
\label{sec:bhsoln}

The following is a black hole solution 
of \eqref{actionLdE}
that asymptotes to the solution written in 
\eqref{4dzerotemp}:
\be 
\label{4dbh}
ds^2_E =e^\sigma\( -Q f{dt^2 \over r^6 } + {d\vec{x}^2 \over r^2} + {dr^2 \over r^2 f}\), ~~B = Q{(1+f) dt \over 2 r^4},~~ e^{2\sigma} =   {r^2 \over Q}\ee
where $f = 1-r^4/r_H^4$. 
The above solution can be obtained (by dimensional reduction) from the following five dimensional solution of Einstein's equation (with negative cosmological constant):
\be ds^{2}_{5} = {Q \over 4 r_{H}^{8}}r^{2}dt^{2} + {d\vec{x}^{2} \over r^{2}} + {(1+f)d\xi dt \over r^{2}} +  {dr^{2} \over f r^{2}} + {r^{2} \over Q} d\xi^{2} \label{slike}\ee 
The scaling symmetry of $Sch$ algebra relates solutions with different values of $Q$ and $r_{H}$. These solutions are also related to each other through the 
lightcone symmetry: $t \rightarrow \lambda t,\xi \rightarrow \lambda^{-1} \xi$. Even though compactification of $\xi$ direction breaks this symmetry, this ``symmetry'' relates a system with chemical potential $\mu$ and temperature $T$ to the system with chemical potential $\mu/\lambda^{2}$ and temperature $T/\lambda$. This transformation along with the conformal Ward identity fixes the form (as a function of chemical potential and temperature) of the free energy \cite{Maldacena:2008wh}.

A curious feature of the lower dimensional solution is the fact that the gauge field is non-vanishing at the horizon. This is not an indication that the solution is irregular. In fact, this solution was obtained from a regular solution in more dimensions. The gauge field obtained by dimensional reduction of the solution in \cite{Maldacena:2008wh, Herzog:2008wg, Adams:2008wt} also has this feature. 

Note also that $t$ is not a time-like direction in (\ref{slike}). However, dimensional reduction of the 
higher-dimensional system results in $t$ becoming a time-like direction. This feature can also be seen in rotating black hole solutions. 

\subsection{Thermodynamics}
\label{sec:boundaryterms}
 
The temperature of the black hole in 
\eqref{4dbh} is 
$T = {\sqrt{Q} \over 2 \pi r_H^3}$ and the entropy density is 
\be s = {\kkk \over  \sqrt{Q} r_{H}} ~.\ee
The energy density, pressure and free energy can be computed from the regularized action. It is possible to regularize the on-shell action and boundary stress tensor with the following boundary counterterms:
\be
\label{boundaryterms}
S_{ct} = \kkk \int_{bdy} d^3 x \sqrt{\gamma}\( -4 e^{- \half \sigma} + e^{3\sigma}\half n^r B^\mu F_{r \mu} \) ~.\ee

Note that 
the second term in (\ref{boundaryterms}) changes the boundary condition
on the gauge field from Dirichlet to Neumann; 
this means that we are in the canonical ensemble (fixed $\rho$).
In its origin as the dimensional reduction 
of the previous $d+3$-dimensional Schr\"odinger solution,
this `Neumannizing' 
term results from the dimensional reduction of the Gibbons-Hawking term.

In the higher-dimensional system, the number density is given by the momentum along the $\xi$ direction. In the lower dimensional system this momentum appears as the charge density of the black hole which is given by
\be \rho = {N\over L_{x}L_{y}} = {\kkk\over L_{x}L_{y}}\int_{bdy}d^{2}x \sqrt{\gamma} n^{r}e^{3\sigma}F_{r}^{t} \sim Q^{-1} \label{density} \ee
Using this we find 
\be-\CF = \CP = \CE = \half {\kkk \rho^{2/3} T^{4/3}} \sim {T^4 \over \mu^2} \label{weirdthermo}\ee
where $\CC$ is a numerical constant.
The chemical potential, read off from
$\partial \CF \over \partial \rho$ , is 
\be \mu \sim {T^{4/3} \over \rho^{1/3} } \quad \mbox{or} \quad \rho \sim {T^{4}\over \mu^{3}}~.
\ee
The form of the thermodynamic quantities are the same as that in \cite{Maldacena:2008wh, Herzog:2008wg, Adams:2008wt}. In the following section, we will present a solution describing a NRCFT with a finite density at zero temperature, which has a non-zero free energy (unlike the $T\to 0$ limit of \eqref{weirdthermo}).

{\bf Note added in v2:} 
The solution \eqref{4dbh} 
and the solution 
\eqref{4dzerotemp} with periodic imaginary time
are saddle points of the same action.  
However, for any $T>0$, the black hole solution \eqref{4dbh}
has a smaller on-shell action and hence its contribution dominates.
Like in planar $AdS$, the would-be Hawking-Page transition is at $T=0$ (where the two solutions
coincide);
unlike in $AdS$, here this does not follow from scale invariance.
However, it has been brought to our attention (by Tom Faulkner)
that the 5d uplift of the solution in this section is in fact isometric to the $AdS_5$ black brane solution;
this explains the similarity in the phase diagram.

\section{A Holographic Mott Insulator?}
\label{sec:mott}

Let us now look at a marginal deformation of the above system;
this will lead to an interesting new family of solutions.
We will do this by adding a massless scalar field $\Psi$ in the bulk. 
The corresponding operator will turn out to be marginally relevant
in the presence of finite density.
Consider the following action with two scalar fields:
\be \label{actionLcut}  S^{E}_{4} = \kkk \int d^{4}x{\(-g_{4}\)}^{1/2}\Big[R_{4} - 2 \Lambda e^{-\sigma}  -{e^{3\sigma}\over4} \(dB\)^2 - {3\over 2} \(\del \sigma \)^2  -\half \(\del \Psi \)^{2}  \Big]
~~.\ee 
The following background is a saddle point of the above action which has asymptotic $Sch$ symmetry:
$$ ds^2_{E}\(\widehat{Sch}\) =e^{\sigma}\( -Q K_{x}^{2}{  dt^2 \over r^6 } + K_{x}{d\vec{x}^2 \over r^2} + {dr^2 \over  r^2 }\)$$ 
\be ~~B = Q{dt \over r^4},~~ e^{2\sigma} =   {r^2 \over Q},~~e^{2\Psi /\sqrt{5}}= 
{1+\varsigma^{2}r^{4}/Q^{2}\over 1-\varsigma^{2}r^{4}/Q^{2}}
\label{solfour}\ee
where $K_{x}^{2}=1 - {\varsigma^{4}r^{8} /Q^{4}}.$
The geometry is cut off at $r=r_{0} = \sqrt{Q/\varsigma}$ where $K_x(r_0)=0$.
There is a curvature singularity at $r = r_0$,
which we resolve below.
$\varsigma$ is a dimensionless parameter describing the source for the operator dual to $\Psi$;
this is a marginally relevant operator whose running produces the dimensional trasmutation 
scale $r_0$ in the solution.
 This solution has non-zero energy, presure, density and free energy\footnote{$\CE = \CP = - \CF \sim \varsigma^2 Q^{-2}$}, but has zero entropy.
 Note that the number density, identical to the calculation of (\ref{density}), is $Q^{-1}$ 
 and the chemical potential is $\varsigma^{2}Q^{-1}$.

The curvature singularity at $r=r_{0}$ 
can be resolved by dimensional oxidation. 
In general, dimensional reduction of a regular solution along a circle action with
degenerate fibers can result in a curvature
singularity in the lower dimensional metric \cite{Gibbons:1994vm}. 
In the next subsection we will show that such a resolution is available here.  

The equations of motion have a symmetry which takes $\Psi \to - \Psi$.
If we identify $\Psi$ with the dilaton field, as we will in the next subsection,
this transformation is an S-duality transformation.
In the solution obtained by the action of this transformation
(this reverses the sign of $\varsigma^2$),
the coupling dual to $\Psi$ is marginally {\it irrelevant}. 
The phase diagram is thus similar to the BCS RG flow, where an attractive/repulsive coupling 
is relevant/irrelevant.
We note, however, that even in our gravity solution for the marginally irrelevant case, 
the flow ends at a finite location in the bulk; perhaps this can 
be attributed to the strong coupling in the dual frame.

\subsection{Lift to eleven dimensions and mass gap}
 
 We begin our journey to a smooth uplift of the solution \eqref{solfour} 
 by noting that the four dimensional action in (\ref{actionLcut}) can be obtained as a consistent truncation of the following five dimensional action:
\be S^{E}_{5} = C_0 \int d^{5}x{\(-g_{5}\)}^{1/2}\Big[R_{5} - 2 \Lambda  -  \half \(\del \Psi \)^{2}  \Big] ~.\label{5D}\ee
In particular, the equations of motion of $S_5^E$ with ansatz 
\be
\label{5dansatz}
ds^2_{5} = e^{- \sigma} ds^2_{E}\(\widehat{Sch}\) + e^{2\sigma} \(d\xi + B\)^2
= ds^2_{E}\(Sch\) + e^{2\sigma} \(d\xi + B\)^2
\ee
are the equations of motion of (\ref{actionLcut});
we have defined 
\be
ds^2_{E}\(Sch\) \equiv 
e^{-\sigma} ds^2_{E}\(\widehat{Sch}\) = 
-Q K_{x}^{2}{  dt^2 \over r^6 } + K_{x}{d\vec{x}^2 \over r^2} + {dr^2 \over  r^2 }~~.
\ee
 In fact, the action in (\ref{actionLdE}) was 
 obtained by dimensional reduction of a five dimensional system 
 related to \eqref{5D} by turning off $\Psi$.

We pause on our path to eleven dimensions to make some comments about
the geometry \eqref{5D}.  
The asymptotics of the 5d metric are precisely AdS with a light-like identification:
\be ds^2 = { 2 d\xi dt + d\vec x^2 + dr^2 \over r^2} ; \ee
this is the  realization of Schr\"odinger symmetry described in \cite{Goldberger:2008vg, Barbon:2008bg}.
Note that with the ansatz in \eqref{5dansatz}, a gauge transformation of the $B$ field
which does not fall off at the boundary has a dramatic effect on the asymptotics.
For example, the transformation $B \to B + \alpha dt $ 
is equivalent in the higher-dimensional description to a redefinition of the $\xi$ coordinate by
$\xi \to \xi + \alpha t$.  
This violates the periodicity of the $\xi$ coordinate and is not an equivalence relation.
Such a transformation is precisely what would be required in order to 
set the gauge field to zero at the IR boundary $B_t(r_0) = 0$.

Some evidence that this solution is not 
the result of Melvinization of a relativistic geometry
is the fact that the free energy is finite
at zero temperature and finite chemical potential;
this is hard to get from a $T\to0$ limit of $F \sim -T^4/\mu^2$.

The five-dimensional action \eqref{5D}
 can in turn be obtained as a consistent truncation of type IIB supergravity \cite{Maldacena:2008wh}. 
 Specifically, the action in (\ref{5D}) can be obtained 
from the consistent truncation of \cite{Maldacena:2008wh}
 by turning off the massive vector
as well as the breathing and squashing modes $u,v$. 
This allows us to lift the solution in (\ref{solfour}) to the following solution of type IIB supergravity:
 $$ds^2_{10} = ds^2_{E}\({Sch}\) + e^{2\sigma} \(d\xi + B\)^2 + ds^2\(S^5\), $$
 \be F_5 =  4 \( \Omega_5 + \star \Omega_5\),~\mbox{and} 
~~\Phi = \Psi  ~~
 \label{typeIIB}\ee
where $\Phi$ is the IIB dilaton.
However, there is a curvature singularity at $r= r_0$ even in this ten dimensional metric. 
The presence of this singularity is related to the non-trivial profile of the dilaton, 
consistent with our interpretation above in terms of dimensional transmutation.
It is convenient to think of the dilaton as the radius of a compact direction in eleven dimensions
\cite{Witten:1995ex}.
 This suggests that the singularity in the 10-D metric can be resolved by lifting it to $M$-theory.
The details of the lift are described in Appendix B.
 After performing the lift, we get the following 11-dimensional solution, which is regular:
  $$ds^2_{11} = e^{-\Psi/6} \Big[ds^2_{E}\({Sch}\) + e^{2\sigma} \(d\xi + B\)^2 + ds^2\(\mathbb{CP}^2\) +  d\chi_1^2 \Big]+ e^{4\Psi/3} d\chi_2^2$$
\be F_{4} = 
 2 J \wedge J 
 + 2  J\wedge d\chi_{1}\wedge d\chi_{2}  \label{11Dsol}\ee
where $J$ is the K\"ahler form on $\mathbb{CP}^{2}$. We can now get two solutions of type IIA from this 11 dimensional solution - (a) by reducing along  $\chi_1$ and (b) by reducing along $\chi_2$. The first
reduction produces a regular metric (with a smoothly shrinking circle) and a constant dilaton, while the second system has a metric with a curvature singularity and non-trivial dilaton profile. 
The second system is related to the type IIB solution in (\ref{typeIIB}) by T-duality. 
The two type-II solutions are related by S-duality. 

Note that in the presence of fermion fields (as in eleven-dimensional supergravity), 
the regularity of the solution 
\eqref{11Dsol} requires antiperiodic boundary conditions around the $\chi_2$ circle
for the fermions, since in the neighborhood of $r_0$, $\chi_2$ is merely an angle
in polar coordinates in $\IR^2$.
This explicitly breaks any supersymmetries.

$\varsigma \equiv Q/r_0^2$ is a dimensionless parameter.  
It can be considered a perturbation of the non-normalizible falloff of $\Psi$,
which from the IIB frame, is the string coupling.  
This encodes a marginally relevant deformation 
of the boundary theory.
In vacuum, it is exactly marginal.  It is driven marginally relevant by the finite density,
and runs strong at $r=r_0$, producing this confining groundstate.


A finite temperature solution can be obtained simply by periodically identifying
the Euclidean time direction in this solution.  
It is not clear that this is the thermodynamically favored solution\footnote
{We note that 
in contrast with the Hawking-Page transition 
\cite{Hawking:1982dh, Witten:1998zw},
in our case a double Wick rotation $ t \to i \chi_2, \chi_2 \to - i t$ does not provide a finite temperature
deconfined solution with the same asymptotics, because of the $dt d\xi $ term in the metric.
The ability to do this previously was a result of Lorentz invariance of the asymptotics.
}.
If it is, then it implies that $e^{ - E_{gap}/T}$ effects
do not appear in observables in this state;
this is consistent with an energy gap of order $N^2$.
At $\varsigma \to 0$, a finite-temperature solution with a horizon
is the one given in Section \ref{sec:bhsoln}.

\subsection{What is a translation-invariant insulator?}

 The fact that the geometry ends smoothly in the IR (at $r_0$) 
strongly suggests that the excitations 
 of this groundstate are gapped \cite{Witten:1998zw}.
More precisely,
regularity requires the boundary condition $\del_{r}\varphi|_{r=r_{0}} = 0$ 
on any smooth 11-dimensional field $\varphi$. 
This {\it real} boundary condition in the IR implies the vanishing
of the spectral density $\Im \vev{\CO \CO}$ of the dual operator $\CO$,
up to a discrete series of delta functions associated with normal modes.
In particular, this applies to the bulk gauge field $B$ 
which couples to the particle number current $j$, and implies
a gap in the spectral density for $j$.  This spectral density determines
the conductivity.

Hence this solution is dual to a system at finite density with a gap for the charged excitations. 
We emphasize that the distinction between this solution
and an ordinary confining groundstate of the dual gauge theory
\cite{Witten:1998zw} is the presence of a nonzero charge density.

Such a thing can be called a Mott insulator.
From the point of view of the dual field theory, 
it is the strong interactions that prevent the charge from moving.
It is certainly not a band insulator or an Anderson insulator -- indeed this system is translationally invariant. 
This raises a thorny point: translation invariance plus finite charge density implies
that the center of mass of the system
will accelerate in an external field, and hence
$\Re \sigma(\omega) \propto \delta(\omega)$ -- 
the DC conductivity is infinite. The system is actually a perfect conductor.


What we mean by calling the system an insulator is 
that we believe it would be an insulator if we pinned it down,
for example 
by a boundary condition.
We have not figured out how to show that the thing is actually an insulator in the above sense.
The answer for the conductivity
$\Re \sigma(\omega) \propto \delta(\omega)$ is not enough:
the clean free Fermi gas also gives this answer,
and obviously that is a metal.
In that case, 
and quite generally \cite{Hartnoll:2007ih}\footnote
{We thank Sean Hartnoll for bringing this argument to our attention.
}, 
adding static impurities just turns the delta function into a transport peak.
In real systems (\ie\ with decent UV behavior) there is a sum rule that says that the spectral weight
from the delta function has to be redistributed somehow upon adding a momentum sink.
A possible concern is that the conclusion 
(\ie\ whether the spectral weight gets redistributed away from $\omega=0$)
might depend on {\it how}
the center of mass mode is frozen.

Even zero compressibility is not enough,
at least in the presence of long-range forces (which presumably the dual field theory has):
the `jellium model' of a metal (in which the lattice of ions is approximated by a
fixed uniform density of background charge)
is incompressible if Coulomb interactions are included, but is also clearly a metal.
Further analysis is required to test our conjecture.
The application of an electric field of finite wavenumber may be 
the simplest approach.

There are several known examples of translation-invariant insulators\footnote
{We thank T.~Grover, S-S.~Lee, T.~Senthil and B.~Swingle for
very helpful conversations on these issues.}.
Quantum Hall states are insulators which preserve continuous translation invariance;
the translation-invariance delta function in
$\sigma(\omega)$ is shifted from zero to the cyclotron frequency because
the (charged) center of mass mode is subjected to a magnetic field
(this is known as Kohn's theorem).
In contrast, the state discussed in this paper is 
not subjected to an external magnetic field.

In strongly-correlated lattice models, the particles can fractionalize
in such a way as to 
produce an integer number of fractionalized particles
per unit cell,
which can then realize an ordinary band insulator.
The arguments of \cite{Kohn, Scalapino, Oshikawa} show that
in a gapped system with a conserved particle number (not spontaneously broken)
at incommensurate filling,
either translation invariance is broken
or the system exhibits groundstate degeneracy on a torus.
All of the examples mentioned above realize the latter option.

For realizing a finite-density insulating state
which preserves continuous (non-magnetic) translations,
it is crucial that the high-energy excitations of 
our system are not particles, but rather CFT excitations.
If the system at the scale of the chemical potential were 
described in terms of charged particles, a state where
the charges were localized would have to (spontaneously) 
break translation invariance,
since the particles have to sit somewhere.
It would be interesting to find a ``slave unparticle" construction of such a state.

\begin{figure}[h!] \begin{center}
\includegraphics[scale=0.5]{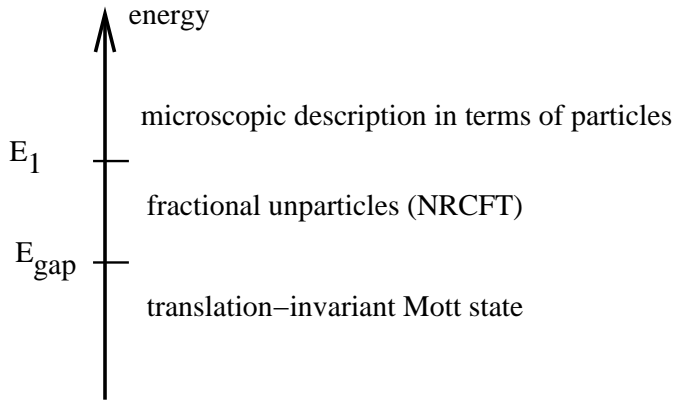}
\caption{\label{fig:scheme} 
A scheme for realization a translation-invariant insulator.
In the gravity dual, we are only addressing physics below the scale $E_1$.
}
\end{center}
\end{figure}

A state in which $\Im \vev{\rho(\omega)\rho( - \omega)}$ 
vanishes below some gap must be incompressible\footnote{We thank D. Son and S. Sachdev for 
emphasizing this to us.  As we learned from Son, this follows 
by comparing the compressibility sum rule and the f-sum rule.}.
Our system, as currently presented, does not have such a gap
(there are zero-energy excitations, at least those associated with translation invariance),
and naively the compressibility is finite.  Indeed it
seems to be a consequence of the scale invariance Ward identity that
$\mu \propto \rho$, which is what we find with an appropriate choice of boundary counterterms.
What this constraint on the compressibility has to say about our proposal for what would happen if one pinned this system down is 
not clear to us at present.
One point to note is that we are forced to study the system at fixed particle number
rather than fixed chemical potential (see \S\ref{sec:boundaryterms}).
In general, an incompressible system ($d\rho/d\mu = 0$) 
does not have a homogeneous groundstate at fixed particle number.
For example, consider a Mott insulator of repulsive bosons on a lattice.
In general only certain values of the number density will admit
homogenous groundstates states (\ie\ those values with integer filling fraction) and at other values the system will phase separate.
The phase separated solutions are compressible.
It may be that the solution we have is describing such a mixed state.
Another possible resolution
is that the nonzero compressibility arises in our system
via neutral modes -- \ie\ in our construction,
it's not clear that all excitations have nonzero particle number.

A final disclaimer about our use of the name ``Mott insulator" is that 
there is no local moment physics in our problem so far.
It would be interesting to include spin degrees of freedom.

\section{Examples with multiple species}
\label{sec:multiplespecies}

In the Schr\"odinger metric of \cite{Son:2008ye, Balasubramanian:2008dm}, the $\xi$-momentum is dual to the particle number $\hat N$
of the dual NRCFT.
Compactifying the $\xi$ direction on a circle of radius $L_\xi$ gives a spectrum of $\hat N$ which 
is just a tower of integer multiples of a fixed mass scale $L_\xi^{-1}$.  
In a system with multiple species of different mass
(for example,
in a pile of atoms consisting
of several species)
the mass operator will have a spectrum which is not just a tower 
of integer multiples of a fixed mass.
Here we would like to find gravity duals with similar spectra.
We can do this by adding more dimensions analogous to $\xi$
\footnote{The discussion in this section was motivated by a question asked by Pavel Kovtun,
and we thank him for discussions.
This question was independently asked by Petr Ho\v{r}ava.}.

We mention in passing that a usefully liberating perspective on the realization of the 
particle number was provided by \cite{Yamada:2008if}: 
in the analog of `global coordinates' discussed there
the particle-number circle
is fibered non-trivially over the $t, \vec x $ directions.

We can solve a reasonable set of equations of motion
with a nice metric with two $\xi$ directions
if we add a second gauge field.
The Lagrangian is just
\be L = R + 2 \Lambda - {1\over 4} F_1^2 - \half m_1^2 A_1^2  - {1\over 4} F_2^2 - \half m_2^2 A_2^2  \ee
with (for $d=2$)
\be 
\label{eq:masses}
m_1^2 = 4z, ~~~~m_2^2 = - 4 (z-2)  
\ee
\be \Lambda = \half\( 26 - 7 z + z^2 \) \ee
(for $z=2$,  $\Lambda = 8$).
The $z$-dependence of $\Lambda$ is a novel development, 
compared to previous Schr\"odinger solutions.

The solution is
\be ds^2 = - r^{-2z} dt^2 + r^{-2} ( - 2 d\xi_+ dt + d\vec x^2 + dr^2 ) + d\xi_-^2 r^{2z - 4} \ee
(the symmetries are discussed below)
with
\begin{equation}
\label{gaugesoln}{ A_1 = \Omega_1 r^{-z} dt }
\end{equation}
$$ A_2 = \Omega_2 r^{z-2} d\xi_- .$$
We believe that it is not possible to source the stress tensor for this metric with a 
single gauge field (which solves its own equations of motion).

Note that the mass-squared of the second
gauge field is negative for many $z$'s of interest ($z>2$).
According to \eqref{eq:masses},
the second gauge field is massless for $z=2$;
we will comment below on some 
subtleties with this case.

\subsection{Symmetries}

This metric is invariant under galilean boosts just like the usual Schrodinger metric,
with no action on $\xi_-$,
\be \xi_+ \to \xi_+ + \vec v \cdot \vec x - \half v^2 t , \xi_- \to \xi_- .\ee
It is scale invariant with
$ \xi_\pm$ both scaling like $length^{2 -z }$:
\be\vec x \to \lambda \vec x  , ~t\to  \lambda^{z} t ,~ r\to \lambda r, ~ \xi_\pm \to \xi_\pm \lambda^{2-z} ~~.\ee
Interestingly, for $z=2$, the $g_{\xi_-\xi_-}$ coefficient is 1.
And, finally, $[K_i, P_j ] = i \delta_{ij} \hat N$ 
with 
\be \hat N = i \del_{\xi_+} ~.\ee

So, if we set $ \xi_\pm = \xi^1 \pm \xi^2 $ and compactify 
\be \xi_1 \simeq \xi_1 + L_1 , ~~ \xi_2 \simeq \xi_2 + L_2 \ee
then the spectrum of $ \hat N$ is 
\be \left\{ {n_1 \over L_1} + { n_2 \over L_2} | n_{1,2} \in \IZ\right\} ;\ee
in particular $L_1 \over L_2 $ needn't be rational.
We can think of $i\del_{\xi_1}$ and $i\del_{\xi_2}$ as the conserved
particle numbers of the individual particle species;
only their sum appears in the Schr\"odinger algebra.

The isometries of this spacetime include
$P_i = i \partial_{x^i} , K^j = i x^j \del_{\xi_+} + i t \partial_{x^j}$
and the Schr\"odinger algebra says:
$ [P_i, K^j] = i\delta_i^j \hat N $
so we have
$\hat N = i \partial_{\xi_+}$.
For $z=2$, there is trivially a special conformal symmetry which
acts on $\xi_+$ in the same way as on $\xi$ in the usual Schr\"odinger 
spacetime, and does not act on $\xi_-$.

It would be interesting to realize a system with arbitrarily many $\xi$-directions (\ie\ species).
Note that the new realization of the Galilean algebra described
in the rest of this paper offers a simple possibility:
one can just introduce a collection of gauge fields
(perhaps coming from some $p$-form reduced on representatives 
of some rank $p-1$ cohomology group of a compactification space 
\eg\ 
as described recently in \cite{Klebanov:2010tj})
and associate them with conserved particle numbers of various species.
The Galilean boost will act by some linear combination of 
the gauge generators; this combination is the total mass
appearing in the Galilean symmetry algebra.

The wave equation is qualitatively the same as in the one-species 
case \cite{Son:2008ye, Balasubramanian:2008dm}.

\subsection{$z\to 2$}

Note that the interesting case $z=2$ is actually quite degenerate here.
The Einstein equations determine the coefficients $\Omega_{1,2}$ in the solutions
for the gauge fields to be (for $d=2$!)
\be \Omega_1^2 = 
2 { z- 1 \over z}~, ~~~~~
 \Omega_2^2 = 
2 { z -1 \over z-2} ~~.\ee
Notice that $\Omega_2$ has a pole at $z=2$. 
But the stress tensor it produces is finite, because both 
the field strength and the mass go to zero as $z\to 2$!  
That is, we must take a scaling limit where $ z\to 2, \Omega_2 \to \infty$ 
holding fixed $ (z-2) \Omega_2^2 $ to which the stress tensor of $A_2$ is proportional.

\section{Comments on the superfluid state}
\label{sec:SF}

The ground state of most assemblies of ultracold atoms,
bosonic or fermionic,
is a superfluid \cite{zwergerreview}.
It is natural to ask whether
the zero-temperature, finite-density solution 
found in \cite{Adams:2008wt}
describes such a state.
That it does not can be seen as follows.
If shifts in the $\xi$-direction correspond to the particle-number symmetry,
then the gravity dual of a superfluid ground state must somehow break translation invariance in the $\xi$ direction
in the IR region of the geometry.
This is because the ground state wave function of a superfluid
is localized
in the space conjugate to the particle number.

A precedent for the required gravity description is the spontaneous breaking of the $U(1)_R$ symmetry
in the Klebanov-Strassler \cite{Klebanov:2000hb}
and Maldacena-Nunez \cite{Maldacena:2000yy}
solutions,
where it is indeed some isometry of the bulk geometry
which is broken (to a discrete subgroup) by the exact solution
in the IR region of the geometry.
A possibility to keep in mind is that the symmetry may be broken
by something other than the metric, \eg\ some other field.

We note that it is not clear that the twisted DLCQ theories,
to which the stringy embeddings of Schr\"odinger spaces found in 
\cite{Maldacena:2008wh, Herzog:2008wg, Adams:2008wt} 
are dual,
indeed have superfluid ground states.
If not, how does the dual field theory avoid breaking the particle number symmetry
at zero temperature?
The fact that the only zero-temperature solution we know \cite{Adams:2008wt} is singular
leaves open the likely possibility that there is a better,
more correct solution with the same leading asymptotics which does describe a superfluid.

In the new realizations of the Schr\"odinger symmetry described in this paper, the 
question of spontaneous breaking of the particle number symmetry 
becomes much more similar to the (well-developed) study of 
holographic superconductors in gravity duals of relativistic CFTs \cite{holographicSC, HSCreviews}.
We note in particular that the system of \S\ref{sec:mott} has a dimensionless parameter
$\varsigma$ which controls the strength of the coupling.
Upon the addition of a charged scalar to the bulk,
we anticipate that varying $\varsigma$ will produce a quantum phase transition 
from the ``Mott" ``insulator" phase described here 
to a superfluid phase \cite{WIP}.

\section{Conclusions}

In this paper we have introduced a new class of gravity duals 
of Galilean-invariant CFTs.  
This requires somewhat novel asymptotics.  In particular, 
the bulk gauge field which represents the particle number symmetry
becomes strongly coupled at the UV boundary.
In the examples constructed by dimensional reduction,
this strong coupling of the gauge field is resolved by the lift;
this is the statement that the $\xi$ direction becomes null 
at the boundary of an asymptotically-Schr\"odinger geometry.
It is an interesting open problem to characterize
the resolution independent of the lift.

In the most interesting new solution we found (described in Section \ref{sec:mott}), 
there was also a singularity at the IR end of the geometry.
This curvature singularity was resolved by a lift to 11-dimensional supergravity;
we emphasize that the shrinking circle in the IR is {\it not} the particle-number direction $\xi$.
It would be interesting to characterize which
singularities of this kind can be resolved (see \eg\ \cite{Gibbons:1994vm, Gubser:2000nd}).
A necessary criterion for a resolution by oxidation is that the 
geometry be conformal to a regular metric.
It would be most useful for our purposes to be able to describe the resolution without
resorting to dimensional oxidation.

Solutions of related Einstein-Maxwell-dilaton systems have been studied recently in
\cite{Gubser:2009qt, notaholographicmottinsulator, Charmousis:2010zz, 
Gursoy:2010jh}, mainly with $AdS$ asymptotics in mind.  
It is possible that the near-infrared solutions studied in these papers can be integrated 
to the asymptotics described here.

Finally, we comment that 
although our goal in this paper was to rid ourselves of the extra dimension $\xi$ conjugate to the particle number,
the 11-dimensional supergravity solution in Section \ref{sec:mott} does indeed include 
such a dimension.  
Further, the regular 4d black hole solutions 
(without a $\xi$ direction)
which we found (in Section \ref{sec:bhsoln} and Appendix \ref{app:bh})
all have equations of state similar to that following from DLCQ \eqref{DLCQthermo}.
It will be of great interest to find black hole solutions,
with the asymptotics described here, which have other equations of state.

\bigskip
\centerline{\bf{Acknowledgements}}

We are grateful to A. Adams, M.~Ammon, J. de Boer, I. Cirac, G. Coss, 
T. Faulkner, 
T. Grover, S. Hartnoll, C. Hoyos, 
S. Kachru, P. Kovtun, 
S-S.~Lee, H. Liu, S. Sachdev, T. Senthil, D. Son, and B. Swingle for discussions and comments.
We thank D. Nickel for collaboration at various stages of this project.
C. Hoyos and D. Son have also considered the dimensional reduction
of the black holes of \cite{Maldacena:2008wh, Herzog:2008wg, Adams:2008wt}.
We would like to thank A. Adams and P. Kovtun for 
emphasizing at an early stage the possible utility of KK reduction along the $\xi$ direction.

This work was supported in part by funds provided by the U.S. Department of Energy
(D.O.E.) under cooperative research agreement DE-FG0205ER41360,
in part by the National Science Foundation under Grant No.\ NSF PHY05-51164,
and in part by the Alfred P. Sloan Foundation.  JM thanks the KITP for hospitality
during the miniprogram on ``Quantum Criticality and the AdS/CFT Correspondence"
where part of this work was done.

\appendix
\section{Black hole solution in a system with two scalars}
\label{app:bh}

In this appendix we present another system in which we have found black holes
with asymptotic Schr\"odinger symmetry.  Its action is rather contrived.

Let us consider the following action
\be
 \label{actionLdnew}  S^{E}_{D} =  \int d^{D}x{\(-g_{D}\)}^{1/2}\Big[R_{D} - 2 \Lambda e^{-\sigma}  -{e^{3\sigma}\over4} \(dB\)^2 - {3\over 2} \(\del \sigma \)^2  -\half \(\del \Psi \)^{2} 
+ V_{2}(\sigma,\Psi) \Big]
\ee
where
\be
 V_{2}(\sigma,\Psi) = \[12 e^{-\sigma}\(\sinh\(\Psi/\sqrt{5}\)\)^3+16\sqrt{Q}/r_{0} \(\tanh\(\Psi/\sqrt{5}\)\)^{-9/4}\(\sinh\(\Psi/\sqrt{5}\)\)^5\]C_{1}^{2}
 \ee
The following background is a saddle point of this action
 \be
ds^2_E =e^{\sigma}\( -Q f K_{x}^{2}{  dt^2 \over r^6 } + K_{x}{d\vec{x}^2 \over r^2} + {dr^2 \over f r^2 }\)$$ $$ ~~B = Q{(1-r^{4}/r_{H}^{4}) dt \over r^4} + B(r_H),~~ e^{2\sigma} =   {r^2 \over Q},~~\Psi =\sqrt{5} \tanh^{-1} \({r^{4}\over r_{0}^{4}}\) \ee
where
\be K_{x}^{2}=1 - {r^{8} /r_{0}^{8}} ~~~{\rm and}~~~
f = 1 - C_{1}^{2} {r^{4} \over \sqrt{r_{0}^{8}-r^{8}}}.\ee
The above system has asymptotic Schr\"odinger symmetry, realized as in Section \ref{sec:sym}. The free energy of this system has the same form as the black hole in the system with one scalar {\it i.e.},
\be \CF  \sim -{T^4 \over \mu^2}.\ee

 \section{Uplifting to M-theory}

In this appendix we exhibit a useful sector of type IIA supergravity as a consistent truncation of 
eleven-dimensional supergravity{\footnote {We will ignore fermions and assume that the Ramond-Ramond
vector $A_{1}$ is turned off.}}, slightly generalizing
the construction in \cite{Duff}.
 Let us consider the following ansatz for the eleven-dimensional line element and the four-form flux:
$$ ds^{2} = {g}_{MN}dx^{M}dx^{N} = {G}_{\mu\nu} dx^{\mu}dx^{\nu} + e^{2\aleph} dz_{10}^{2}$$
\be \tilde{F}_{4} = F_{4} + H_{3}\wedge dz_{10} \ee  
where ${g}$ is the eleven-dimensional metric, ${G}$ is the ten-dimensional metric, $\tilde F_{4}$ is 
the eleven-dimensional four-form flux, $F_{4}$ and $H_{3}$ are the ten-dimensional four-form and three-form flux. With this ansatz, the Bianchi identity becomes
\be d\tilde{F}_{4} = 0 \quad  \Leftrightarrow \quad dF_{4} = 0\quad \mbox{and} \quad dH_{3} = 0 .\ee
The equation of motion for the eleven-dimensional four-form field strength can be written as
\be d\star\tilde{F}_{4} = {1\over 4} \tilde{F}_{4}\wedge \tilde{F}_{4}  \Leftrightarrow d\(e^{\aleph}\star{F}_{4}\) = 
\half H_{3} \wedge F_{4} \quad \mbox{and} \quad d\(e^{-\aleph}\star{H}_{3}\) = {1\over 4} F_{4} \wedge F_{4} ~.\ee
The components of eleven-dimensional Ricci tensor $(\tilde{R}_{\mu\nu})$ are given by
\be  \tilde{R}_{\mu\nu} = R_{\mu\nu} - \nabla_{\mu}\nabla_{\nu}\aleph - \nabla_{\mu}\aleph\nabla_{\nu}\aleph\ee
\be \tilde{R}_{\mu 10} = 0 \ee
\be \tilde{R}_{10 10} = \(\nabla_{\mu} \nabla^{\mu} \aleph + \nabla_\mu{\aleph}\nabla^{\mu}\aleph\) ~~\ee
where $R_{\mu\nu}$ is the ten-dimensional Ricci scalar. After some algebra, the eleven-dimensional Einstein equations can be written as
\be \(R_{\mu\nu} - \half  G_{\mu \nu} R\) - \nabla_{\mu}\nabla_{\nu}\aleph - \nabla_{\mu}\aleph\nabla_{\nu}\aleph + \(\nabla_{\mu} \nabla^{\mu} \aleph + \nabla_\mu{\aleph}\nabla^{\mu}\aleph\) G_{\mu \nu} = T^{F}_{\mu \nu} + T^{H}_{\mu \nu}\ee
\be R - {1\over 48} F_{4}^{2}  + {1\over 12} e^{-2\aleph}H_{3}^{2} = 0~~.\ee
The above equations can be obtained from the following action\footnote
{We have made use of the following formula 
\be {1\over \sqrt{g} }{\delta \over \delta g_{ab}} \int d^{d}x \sqrt{g}X R = X\(R_{ab} - \half R g_{ab}\) - \nabla_{a}\nabla_{b}X + g_{ab} \nabla_{c}\nabla^{c}X~~ .\ee}
\be S_{10} = \int d^{10}x\sqrt{G}e^{\aleph}\(R  - {1\over 48}F_{4}^{2} -{e^{-2\aleph}\over 12} H_{3}^{2}\) + {1\over 2} \int B_{2} \wedge F_{4} \wedge F_{4}\label{TypeIIAp}\ee
where $H_{3} = dB_{2}$. 
Let us redefine $\tilde{G}_{\mu \nu} = e^{-\aleph}G_{\mu \nu}$ and $\aleph = 2\Phi/3$. In terms of the redefined variables, the action in (\ref{TypeIIAp}) can be written as
\be S_{10} = \int d^{10}x\sqrt{\tilde{G}}e^{-2\Phi}\(\tilde{R}  + 4 \(\del \Phi\)^{2}- {e^{2\Phi}\over 48}F_{4}^{2} -{1\over 12} H_{3}^{2}\) + {1\over 2} \int B_{2} \wedge F_{4} \wedge F_{4} ~~. \label{TypeIIA}\ee
This action is the bosonic part of the type IIA supergravity action (with $A_{1}$ turned off) in string frame. Any solution of the ten dimensional action in (\ref{TypeIIA}) can be oxidized to give a solution of 
eleven-dimensional supergravity. The 10-D Einstein frame metric metric is related to the string frame metric $\tilde G$  through the following Weyl transformation
$ g_{E} = e^{\Phi/2}\tilde{G}.$


\begin{thebibliography}{99}



\bibitem{Son:2008ye}
  D.~T.~Son,
  ``Toward an AdS/cold atoms correspondence: a geometric realization of the
  Schroedinger symmetry,''
  Phys.\ Rev.\  D {\bf 78}, 046003 (2008)
  [arXiv:0804.3972 [hep-th]].
  
\bibitem{Balasubramanian:2008dm}
  K.~Balasubramanian and J.~McGreevy,
  ``Gravity duals for non-relativistic CFTs,''
  Phys.\ Rev.\ Lett.\  {\bf 101}, 061601 (2008)
  [arXiv:0804.4053 [hep-th]].
  
\bibitem{Mehen:1999nd}
  T.~Mehen, I.~W.~Stewart and M.~B.~Wise,
  ``Conformal invariance for non-relativistic field theory,''
  Phys.\ Lett.\  B {\bf 474}, 145 (2000)
  [arXiv:hep-th/9910025].

\bibitem{Nishida:2007pj}
  Y.~Nishida and D.~T.~Son,
``Nonrelativistic conformal field theories,''
  Phys.\ Rev.\  D {\bf 76}, 086004 (2007)
  [arXiv:0706.3746 [hep-th]].


\bibitem{Maldacena:2008wh}
  J.~Maldacena, D.~Martelli and Y.~Tachikawa,
  ``Comments on string theory backgrounds with non-relativistic conformal symmetry,''
JHEP {\bf 0810} (2008) 072
[arXiv:0807.1100 [hep-th]].  
  
\bibitem{Herzog:2008wg}
  C.~P.~Herzog, M.~Rangamani and S.~F.~Ross,
  ``Heating up Galilean holography,''
JHEP {\bf 0811} (2008) 080
  [arXiv:0807.1099 [hep-th]]. 

\bibitem{Adams:2008wt}
  A.~Adams, K.~Balasubramanian and J.~McGreevy,
``Hot Spacetimes for Cold Atoms,''
  JHEP {\bf 0811} (2008) 059 
  [arXiv:0807.1111 [hep-th]].




\bibitem{Kovtun:2008qy}
  P.~Kovtun and D.~Nickel,
  ``Black holes and non-relativistic quantum systems,''
  Phys.\ Rev.\ Lett.\  {\bf 102}, 011602 (2009)
  [arXiv:0809.2020 [hep-th]].

\bibitem{Hartnoll:2008rs}
  S.~A.~Hartnoll and K.~Yoshida,
  ``Families of IIB duals for nonrelativistic CFTs,''
  JHEP {\bf 0812}, 071 (2008)
  [arXiv:0810.0298 [hep-th]].


\bibitem{Colgain:2009wm}
  E.~O.~Colgain and H.~Yavartanoo,
  JHEP {\bf 0909}, 002 (2009)
  [arXiv:0904.0588 [hep-th]];
  N.~Bobev, A.~Kundu and K.~Pilch,
  JHEP {\bf 0907}, 107 (2009)
  [arXiv:0905.0673 [hep-th]];
  A.~Donos and J.~P.~Gauntlett,
  JHEP {\bf 0907}, 042 (2009)
  [arXiv:0905.1098 [hep-th]];
  H.~Ooguri and C.~S.~Park,
  Nucl.\ Phys.\  B {\bf 824}, 136 (2010)
  [arXiv:0905.1954 [hep-th]];
  E.~O' Colgain, O.~Varela and H.~Yavartanoo,
  JHEP {\bf 0907}, 081 (2009)
  [arXiv:0906.0261 [hep-th]];
  A.~Donos and J.~P.~Gauntlett,
  JHEP {\bf 0910}, 073 (2009)
  [arXiv:0907.1761 [hep-th]];
\bibitem{Jeong:2009aa}
  J.~Jeong, H.~C.~Kim, S.~Lee, E.~O.~Colgain and H.~Yavartanoo,
  JHEP {\bf 1003}, 034 (2010)
  [arXiv:0911.5281 [hep-th]].

    
 \bibitem{bateman}
 H.~H.~Bateman,
{\it The mathematical analysis of electric and optical wave motion},
 reprinted by Dover (1955).
 
\bibitem{Duval}
  C.~Duval, G.~Burdet, H.~P.~Kunzle and M.~Perrin,
 ``Bargmann Structures And Newton-Cartan Theory,''
  Phys.\ Rev.\  D {\bf 31}, 1841 (1985).

\bibitem{Horvathy}
  C.~Duval, G.~W.~Gibbons and P.~Horvathy,
  ``Celestial Mechanics, Conformal Structures, and Gravitational Waves,''
  Phys.\ Rev.\  D {\bf 43}, 3907 (1991)
  [arXiv:hep-th/0512188].

%
%
  
\bibitem{BrittoPacumio:1999sn}
  R.~Britto-Pacumio, A.~Strominger and A.~Volovich,
  ``Holography for coset spaces,''
  JHEP {\bf 9911}, 013 (1999)
  [arXiv:hep-th/9905211].
  
\bibitem{Taylor:2000xf}
  M.~Taylor,
``Holography for degenerate boundaries,''
  arXiv:hep-th/0001177.

\bibitem{Yamada:2008if}
  D.~Yamada,
  ``Thermodynamics of Black Holes in Schroedinger Space,''
  Class.\ Quant.\ Grav.\  {\bf 26}, 075006 (2009)
  [arXiv:0809.4928 [hep-th]].
  

\bibitem{Colgain:2009wm}
  E.~O.~Colgain and H.~Yavartanoo,
  JHEP {\bf 0909}, 002 (2009)
  [arXiv:0904.0588 [hep-th]];
  N.~Bobev, A.~Kundu and K.~Pilch,
  JHEP {\bf 0907}, 107 (2009)
  [arXiv:0905.0673 [hep-th]];
  A.~Donos and J.~P.~Gauntlett,
  JHEP {\bf 0907}, 042 (2009)
  [arXiv:0905.1098 [hep-th]];
  H.~Ooguri and C.~S.~Park,
  Nucl.\ Phys.\  B {\bf 824}, 136 (2010)
  [arXiv:0905.1954 [hep-th]];
  E.~O' Colgain, O.~Varela and H.~Yavartanoo,
  JHEP {\bf 0907}, 081 (2009)
  [arXiv:0906.0261 [hep-th]];
  A.~Donos and J.~P.~Gauntlett,
  JHEP {\bf 0910}, 073 (2009)
  [arXiv:0907.1761 [hep-th]];
\bibitem{Jeong:2009aa}
  J.~Jeong, H.~C.~Kim, S.~Lee, E.~O.~Colgain and H.~Yavartanoo,
  JHEP {\bf 1003}, 034 (2010)
  [arXiv:0911.5281 [hep-th]].
  


\bibitem{fuertesDLCQ}
  J.~L.~F.~Barbon and C.~A.~Fuertes,
  ``Ideal gas matching for thermal Galilean holography,''
 Phys.\ Rev.\  D {\bf 80}, 026006 (2009)
  [arXiv:0903.4452 [hep-th]].  


\bibitem{nullmelvin}
  A.~Bergman, K.~Dasgupta, O.~J.~Ganor, J.~L.~Karczmarek and G.~Rajesh,
  ``Nonlocal field theories and their gravity duals,''
  Phys.\ Rev.\  D {\bf 65}, 066005 (2002)
  [arXiv:hep-th/0103090];
   M.~Alishahiha and O.~J.~Ganor,
  ``Twisted backgrounds, pp-waves and nonlocal field theories,''
  JHEP {\bf 0303}, 006 (2003)
  [arXiv:hep-th/0301080].
  
  
\bibitem{Hellerman:1997yu}
  S.~Hellerman and J.~Polchinski,
  ``Compactification in the lightlike limit,''
  Phys.\ Rev.\  D {\bf 59}, 125002 (1999)
  [arXiv:hep-th/9711037].
  
  
%
%
%

  
\bibitem{inonu}  
E. In\"onu and E. P. Wigner, ``Representations of the Galilei group,'' Nuovo Cimento {\bf 9}, 
705 (1952). 


\bibitem{Hartnoll:2009qx}
  S.~A.~Hartnoll,
  ``Quantum Critical Dynamics from Black Holes,''
  arXiv:0909.3553 [cond-mat.str-el];
  ``Lectures on holographic methods for condensed matter physics,''
  Class.\ Quant.\ Grav.\  {\bf 26}, 224002 (2009)
  [arXiv:0903.3246 [hep-th]].

\bibitem{McGreevy:2009xe}
  J.~McGreevy,
  ``Holographic duality with a view toward many-body physics,''
  arXiv:0909.0518 [hep-th], to appear in AHEP.

\bibitem{wittenbagger}
  E.~Witten and J.~Bagger,
  ``Quantization Of Newton's Constant In Certain Supergravity Theories,''
  Phys.\ Lett.\  B {\bf 115}, 202 (1982).

\bibitem{AlvarezGaume:1983ig}
  L.~Alvarez-Gaume and E.~Witten,
  ``Gravitational Anomalies,''
  Nucl.\ Phys.\  B {\bf 234}, 269 (1984).

\bibitem{hellerman}
  S.~Hellerman,
  ``A Universal Inequality for CFT and Quantum Gravity,''
  arXiv:0902.2790 [hep-th].


\bibitem{Nishioka:2009zj}
  T.~Nishioka, S.~Ryu and T.~Takayanagi,
  JHEP {\bf 1003}, 131 (2010)
  [arXiv:0911.0962 [hep-th]].



%
%

  
  
\bibitem{Goldberger:2008vg}
  W.~D.~Goldberger,
  ``AdS/CFT duality for non-relativistic field theory,''
 JHEP {\bf 0903}, 069 (2009)
  [arXiv:0806.2867 [hep-th]].

\bibitem{Barbon:2008bg}
  J.~L.~B.~Barbon and C.~A.~Fuertes,
  ``On the spectrum of nonrelativistic AdS/CFT,''
 JHEP {\bf 0809}, 030 (2008)
  [arXiv:0806.3244 [hep-th]].
  
    
\bibitem{Kachru:2008yh}
  S.~Kachru, X.~Liu and M.~Mulligan,
  ``Gravity Duals of Lifshitz-like Fixed Points,''
  Phys.\ Rev.\  D {\bf 78}, 106005 (2008)
  [arXiv:0808.1725 [hep-th]].

  
\bibitem{Gibbons:1994vm}
G.~W.~Gibbons, G.~T.~Horowitz and P.~K.~Townsend,
``Higher Dimensional Resolution of Dilatonic Black Hole Singularities,''
Class.\ Quant.\ Grav.\  {\bf 12} (1995) 297
[arXiv:hep-th/9410073].


\bibitem{Gubser:2000nd}
S.~S.~Gubser,
``Curvature Singularities: the Good, the Bad, and the Naked,''
Adv.\ Theor.\ Math.\ Phys.\  {\bf 4} (2000) 679
[arXiv:hep-th/0002160].



\bibitem{Witten:1995ex}
  E.~Witten,
  ``String theory dynamics in various dimensions,''
  Nucl.\ Phys.\  B {\bf 443}, 85 (1995)
  [arXiv:hep-th/9503124].
 
    
\bibitem{Hawking:1982dh}
  S.~W.~Hawking and D.~N.~Page,
``Thermodynamics Of Black Holes In Anti-De Sitter Space,''
  Commun.\ Math.\ Phys.\  {\bf 87}, 577 (1983).
 
   
  \bibitem{Witten:1998zw}
  E.~Witten,
  ``Anti-de Sitter space, thermal phase transition, and confinement in  gauge
  theories,''
  Adv.\ Theor.\ Math.\ Phys.\  {\bf 2}, 505 (1998)
  [arXiv:hep-th/9803131].
  
\bibitem{Hartnoll:2007ih}
  S.~A.~Hartnoll, P.~K.~Kovtun, M.~Muller and S.~Sachdev,
  ``Theory of the Nernst effect near quantum phase transitions in condensed
  matter, and in dyonic black holes,''
  Phys.\ Rev.\  B {\bf 76}, 144502 (2007)
  [arXiv:0706.3215 [cond-mat.str-el]].
  
  \bibitem{Kohn}
  W.~Kohn, ``Theory of the Insulating State," Phys.\ Rev.\ {\bf 133},  A171 (1964).
  
  
  \bibitem{Scalapino}
  D.~J.~Scalapino, S.~R.~White, S.~C.~Zhang, ``Superfluid Density and the Drude Weight of the Hubbard Model,''
  Phys.\ Rev.\ Lett.\ {\bf 68}, 2830 (1992).
  
  \bibitem{Oshikawa}
  M.~Oshikawa, ``Insulator, conductor and commensurability: a topological approach,"
  Phys. Rev. Lett. 90, 236401 (2003); Phys. Rev. Lett. 91, 109901(E) (2003)
[arXiv:cond-mat/0301338v2 [cond-mat.str-el]].
  
  

  
\bibitem{Klebanov:2010tj}
  I.~R.~Klebanov, S.~S.~Pufu and T.~Tesileanu,
  ``Membranes with Topological Charge and AdS4/CFT3 Correspondence,''
  arXiv:1004.0413 [hep-th].
  
  
  

\bibitem{zwergerreview}
I.~Bloch, J.~Dalibard, W.~Zwerger,
``Many-Body Physics with Ultracold Gases,"
Rev.\ Mod.\ Phys.\ {\bf 80}, 885 (2008)
[arXiv:0704.3011v2 [cond-mat.other]].


  
\bibitem{Klebanov:2000hb}
I.~R.~Klebanov and M.~J.~Strassler,
``Supergravity and a Confining Gauge Theory: Duality Cascades and   $\chi$SB-Resolution of Naked Singularities,''
JHEP {\bf 0008} (2000) 052
[arXiv:hep-th/0007191].


\bibitem{Maldacena:2000yy}
J.~M.~Maldacena and C.~Nunez,
``Towards the Large $N$ Limit of Pure ${\mathcal{N}}\!=1$ Super Yang Mills,''
Phys.\ Rev.\ Lett.\  {\bf 86} (2001) 588
[arXiv:hep-th/0008001].


  
  \bibitem{holographicSC}
  S.~S.~Gubser,
 ``Breaking an Abelian gauge symmetry near a black hole horizon,''
  Phys.\ Rev.\  D {\bf 78}, 065034 (2008)
  [arXiv:0801.2977 [hep-th]];
  S.~A.~Hartnoll, C.~P.~Herzog and G.~T.~Horowitz,
  ``Building a Holographic Superconductor,''
  Phys.\ Rev.\ Lett.\  {\bf 101}, 031601 (2008)
  [arXiv:0803.3295 [hep-th]];
{\it ibid}, ``Holographic Superconductors,''
  JHEP {\bf 0812}, 015 (2008)
  [arXiv:0810.1563 [hep-th]].
  
\bibitem{HSCreviews}
G.~T.~Horowitz, ``Introduction to Holographic Superconductors,"
arXiv:1002.1722 [hep-th];
  C.~P.~Herzog,
  ``Lectures on Holographic Superfluidity and Superconductivity,''
  arXiv:0904.1975 [hep-th].
  
  \bibitem{WIP}
  Work in progress.
  
\bibitem{Gubser:2009qt}
S.~S.~Gubser and F.~D.~Rocha,
``Peculiar Properties of a Charged Dilatonic Black Hole in $\mathrm{AdS}_5$,''
  Phys.\ Rev.\  D {\bf 81}, 046001 (2010)
  [arXiv:0911.2898 [hep-th]].

\bibitem{notaholographicmottinsulator}
K.~Goldstein, S.~Kachru, S.~Prakash and S.~P.~Trivedi,
``Holography of Charged Dilaton Black Holes,''
arXiv:0911.3586 [hep-th].


\bibitem{Charmousis:2010zz}
  C.~Charmousis, B.~Gouteraux, B.~S.~Kim, E.~Kiritsis and R.~Meyer,
  ``Effective Holographic Theories for low-temperature condensed matter
  systems,''
  arXiv:1005.4690 [hep-th].
  
\bibitem{Gursoy:2010jh}
  U.~Gursoy,
  ``Continuous Hawking-Page transitions in Einstein-scalar gravity,''
  arXiv:1007.0500 [hep-th].

  


  
  \bibitem{Duff}
M.~J.~Duff, H.~L\"u, C.~N.~Pope
``$AdS_{5}\times S^{5}$ untwisted''
[arXiv:9803061[hep-th]].
  

  
\end{thebibliography}
\end{document}